\documentstyle[preprint,aps,epsf]{revtex}
\newif\iftightenlines\tightenlinesfalse
\tightenlines\tightenlinestrue

\def\to{\rightarrow}

\def\te{\tilde e}

\def\tu{\tilde u}

\def\tb{\tilde b}

\def\td{\tilde d}

\def\tst{\tilde t}
\def\ttau{\tilde \tau}

\def\tg{\tilde g}
\def\tnu{\tilde\nu}
\def\tell{\tilde\ell}

\def\tw{\widetilde\chi^\pm}

\def\tz{\widetilde\chi^0}
\begin{document}
\draft
\preprint{\vbox{\baselineskip=14pt%
   \rightline{FSU-HEP-010701}
}}
\title{SUPERSYMMETRIC $SO(10)$ GUT MODELS WITH \\
YUKAWA UNIFICATION AND A POSITIVE $\mu$ TERM}
\author{Howard Baer$^1$ and Javier Ferrandis$^{2}$}
\address{
$^1$Department of Physics,
Florida State University,
Tallahassee, FL 32306 USA
}
\address{
$^2$Institut de F\'\i sica Corpuscular (C.S.I.C.), 
Ap. 22085, Val\`encia 46071, Spain}
\date{\today}
\maketitle
\begin{abstract}

Supersymmetric grand unified models based on $SO(10)$ gauge symmetry 
have many desireable features, including the unification of
Yukawa couplings. 
Including $D$-term contributions to scalar masses arising 
from the breakdown of 
$SO(10)$, Yukawa coupling unification only to 30\% can be achieved in
models with a positive superpotential Higgs mass.
The superparticle mass spectrum is highly constrained,
and yields relatively light top squarks and charginos.  
Surprisingly,
the pattern of GUT scale soft SUSY breaking masses are close to those
found in the context of inverted hierarchy models. 
Our analysis supports the idea that the low
energy MSSM parameter space is an approximate $SO(10)$ inspired
fixed point. 
\end{abstract}

\medskip



Supersymmetric grand unified theories (SUSYGUTS) 
based on the gauge group $SO(10)$
are especially attractive\cite{review}. 
Not only do they unify the three forces 
of the Standard Model (SM), but they unify the matter content of
each generation into
a single 16 dimensional irreducible representation of $SO(10)$: 
$\hat{\psi}({\bf 16})$. The $\hat{\psi}({\bf 16})$ includes not 
only the matter
superfields of the minimal supersymmetric standard model (MSSM), 
but also a gauge singlet superfield $\hat{N}^c$ which includes
a right handed neutrino. The gauge singlet superpotential mass
term $M_N$ can be of order $M_{GUT}$, and leads to sub-eV scale
masses for left handed neutrinos (in accord with SuperK results
on atmospheric neutrino oscillations), while the right handed
neutrinos decouple via the well-known see-saw mechanism\cite{seesaw}.
In the simplest models, the two Higgs doublet superfields of the 
MSSM reside in a single
ten dimensional representation $\hat{\phi}({\bf 10})$.
Then $SO(10)$ SUSYGUT models contain a superpotential
interaction term
\begin{equation}
\hat{f}\ni f\hat{\psi}\hat{\psi}\phi+\cdots 
\label{eq:1}
\end{equation}
where $f$ is the Yukawa coupling 
which leads to masses for quarks and leptons (at this stage we neglect 
intergenerational mixing effects).
Thus, the Yukawa couplings of each generation are assumed to be unified 
above the GUT scale. More sophisticated treatments of the Yukawa
matrices can allow for predictions of all SM fermion masses and 
mixing angles in terms of just a few parameters\cite{fmasses}. 
Here we will focus only on third generation Yukawa
couplings, since they will be large, and can have a substantial impact on
the spectrum of superpartners.

In much the same way that the three gauge couplings of the MSSM can be
extrapolated from their weak scales values to their GUT scale values
using renormalization group (RG) evolution, so too can the 
Yukawa couplings be evolved from the weak to the GUT scale to test
models with Yukawa coupling unification. In the bottom-up 
approach used in ISAJET v7.58\cite{isajet}, 
we begin with the weak scale values of
$m_b$, $m_t$ and $m_\tau$ in the $\overline{DR}$ regularization scheme:
$m_\tau(M_Z)=1.7463$ GeV, $m_b(M_Z)=2.92$ GeV, and $m_t(m_t)=163.4$ GeV.
We calculate the corresponding Yukawa couplings, and
evolve both Yukawa and gauge couplings to higher energies using 
two-loop RG equations (RGEs)\cite{mv}. Once $M_{GUT}$ is determined
(by the point at which the $SU(2)$ and $U(1)$ gauge couplings meet),
we evolve gauge couplings, Yukawa couplings and soft SUSY breaking (SSB) mass
terms to the weak scale $M_{weak}$, 
where  electroweak symmetry is broken radiatively (REWSB), and 
where the entire SUSY particle mass spectrum can be 
calculated. 
At this stage, the Yukawa couplings can be updated to 
include SM and MSSM loop corrections\cite{yuk,bmpz}, 
and the RGE process is iterated
until a convergent spectrum of SUSY particle masses is obtained.
In this way, the GUT scale values of the Yukawa couplings depend on the
SUSY particle mass spectrum.

In previous reports\cite{others,so10,so10_2}, 
unification of Yukawa couplings was investigated 
within the context of $SO(10)$ SUSYGUT models. It is well known that
in the mSUGRA model, with universal SSB masses at the GUT scale, 
a high degree of Yukawa coupling unification only occurs for
negative values of the superpotential Higgs mass term $\mu$, and
for values of the ratio of Higgs field vevs $\tan\beta\equiv {v_u\over v_d}
\sim 50$\cite{bmpz,so10}. 
For such high values of $\tan\beta$, the scalar potential
no longer has the appropriate form at the weak scale to accommodate
REWSB. Generally, the SSB down Higgs
mass squared $m_{H_d}^2$ gets driven to negative values before the
up Higgs mass squared $m_{H_u}^2$. However, even if scalar masses
are universal above the GUT scale, they will receive 
$D$-term mass contributions at the GUT scale arising from the breakdown of
$SO(10)$ gauge symmetry. Thus, scalar masses are shifted by an amount
\begin{eqnarray*}
m_Q^2 &=&m_E^2=m_U^2=m_{16}^2+M_D^2 \\
m_D^2 &=&m_L^2=m_{16}^2-3M_D^2 \\
m_N^2&=&m_{16}^2+5M_D^2 \\
m_{H_{u,d}}^2&=&m_{10}^2\mp 2M_D^2
\end{eqnarray*}
where $M_D^2$ parametrizes our ignorance of the exact breakdown mechanism
for $SO(10)$, and can have either positive or negative values
of order $M_{weak}^2$.
Thus, the model is characterized by the following free parameters:
\begin{equation}
m_{16},\ m_{10},\ M_D^2,\ m_{1/2},\ A_0,\ \tan\beta\ {\rm and}\ sign(\mu ).
\end{equation}
The value of $\tan\beta$ will be restricted by the requirement of
Yukawa coupling unification to be
close to $\sim 50$. In this model, for positive values of $M_D^2$, the
GUT scale values of $m_{H_u}^2$ and $m_{H_d}^2$ are split, and
$m_{H_u}^2$ gets a head start on running towards negative values.
For a sufficiently large value of $M_D^2$, REWSB can be recovered, 
and viable supersymmetric models with a high degree of third 
generation Yukawa coupling can be generated\cite{so10}. 

In previous works,
model parameter space was mapped out under the restriction of
GUT scale Yukawa unification to 5\%\cite{so10}, and implications of
the model for cosmological neutralino relic density, direct 
detection of dark matter, radiative decays $b\to s\gamma$ and collider
searches were determined\cite{so10_2}. 
A favorable relic density was obtained over much of
model parameter space owing to $s$-channel neutralino pair 
annihilation via the very wide $A$ and $H$ Higgs poles at large
$\tan\beta$\cite{relic}. 
However, the decay width for $b\to s\gamma$ was found to be
very large and generally in disagreement with experimental limits 
unless one entered the decoupling regime, where model
parameters began becoming unnatural. The large $b\to s \gamma$
branching fraction at large $\tan\beta$ and $\mu <0$ is well known\cite{bsg}. 
However, for $\mu >0$, the $b\to s\gamma$ branching ratio can be in accord
with experimental limits at large $\tan\beta$. 

In addition, the recent 
measurement by the E821 experiment\cite{e821} 
of the muon anomalous magnetic moment $a_\mu ={(g-2)/2}$ was
found to deviate from SM predictions\cite{cm} by $2.6\sigma$.
If the deviation is interpreted in terms of supersymmetric models, then it
disfavors models with $\mu <0$, assuming positive SSB gaugino 
masses\cite{bbft}.
For these reasons, it seemed prudent to re-examine Yukawa unification 
for positive values of $\mu$, relaxing the ad-hoc $5\%$
criteria used in Refs. \cite{so10,so10_2}.

Our procedure is as follows. We generate random samples of model
parameters
\begin{eqnarray*}
0&<&m_{16}<2000\ {\rm GeV},\\
0&<&m_{10}<2500\ {\rm GeV},\\
0&<&m_{1/2}<1000\ {\rm GeV},\\
-1000^2&<&M_D^2<+1000^2\ {\rm GeV}^2,\\
45&<&\tan\beta <55, \\
-5000&<& A_0<5000\ {\rm GeV}\ {\rm and} \\
\mu &>&0\ .
\end{eqnarray*}
We then calculate the non-universal scalar masses according to
formulae given above, and enter the parameters into the 
computer program ISASUGRA. 
ISASUGRA is a part of the ISAJET package\cite{isajet}
which calculates an iterative solution to the 26 coupled 
RGEs of the MSSM. We required unification of third generation Yukawa
couplings at the GUT scale to 45\%. 
Our requirement of 45\% unification is determined by defining the 
variables $r_{b\tau}$, $r_{tb}$ and $r_{t\tau}$, where for instance
$r_{b\tau}= max(f_b/f_\tau,f_\tau /f_b )$. We then require
$R= max(r_{b\tau},r_{tb},r_{t\tau}) <1.45$. 
We were able to find many solutions fulfilling the above criteria, with 
the best overall unification achieved for $R=1.28$, or Yukawa 
coupling unification to 28\%. The top and tau Yukawa couplings could
frequently unify to very high precision, while $f_b/f_t$ and
$f_b/f_\tau$ could unify to $\sim 0.7-0.8$. 
The Yukawa couplings at the GUT scale, $f_i(M_{GUT})$, 
should differ from the unified Yukawa, $f_{GUT}$, 
due to threshold corrections, 
$f_i(M_{GUT})=f_{GUT} (1+\epsilon_i)$. 
It would be difficult to
explain our $30-40\%$ deviation from perfect Yukawa unification
by GUT scale threshold corrections in simple SUSYGUT models\cite{bwright}, 
where threshold corrections are expected of a few percent.
More complicated models may be needed to explain the large threshold
corrections needed for $SO(10)$ SUSYGUT models 
with $\mu >0$\cite{Bagger:1995bw}.

The parameter space regions with Yukawa coupling unification to 45\%
are shown in Fig. \ref{fig1}. Dots represent allowed solutions, while
crosses denote solutions in violation of the experimental
constraints $m_{\tw_1}<100$ GeV, $m_h<113$ GeV and $m_{\ttau_1}<73$ GeV. 
The values of $\tan\beta$ generated were within the narrow
range of $\tan\beta\sim 46-48$, typically somewhat lower than results 
assuming $\mu <0$. 
In frame {\it a}), we show models in the
$m_{16}\ vs.\ m_{1/2}$ plane. The values of $m_{16}$ and $m_{10}$
typically lie beyond 1 TeV and
cluster about the line $m_{10}=\sqrt{2} m_{16}$.

In frame {\it b}), we show solutions in the $m_{16}\ vs.\ M_D$ plane.
In this case, we find $M_D$ restricted to values of $-0.4$ to $+0.4$ TeV.
Solutions can be obtained with $M_D\simeq 0$, which
brings us back to the mSUGRA model. This contrasts with the
$\mu <0$ case where $5\%$ unification required strictly positive
$D$-terms. 

In frame {\it c}), we show the $m_{16}\ vs.\ A_0$ parameter plane.
Surprisingly, the best Yukawa unified solutions are found only for
large negative values of the $A_0$ parameter, and cluster about the
line $A_0=-2m_{16}$.
Finally, in frame {\it d}), we show the $m_{16}\ vs.\ m_{1/2}$ plane, 
and find model solutions occuring for $m_{1/2}\sim 0.1-0.9$ TeV,
with $m_{16}$ always greater than $m_{1/2}$. 

It is particularly intriguing that model solutions with 
the best Yukawa
coupling unification cluster about model parameters with
\begin{equation}
A_0^2=2m_{10}^2=4m_{16}^2 .
\label{eq:2}
\end{equation}
These particular boundary conditions were found by Bagger {\it et al.}
from a rather different approach\cite{bfpz}, by looking 
analytically for fixed point behavior in third generation SSB
scalar masses which would give rise to SUSY models with a
radiatively driven inverted scalar mass hierarchy (RIMH).
In these models, one may begin with GUT scale scalar masses beyond the TeV
scale. RG evolution drives third generation and Higgs scalar masses 
towards zero, while scalars of the first two generations remain
beyond the TeV range. In this way, multi-TeV first and second generation
scalar masses act to suppress SUSY flavor and CP violating processes, 
while still fulfilling conditions of naturalness, which mainly apply to
the sub-TeV third generation and Higgs scalar SSB masses.
The RIMH mechanism is viable for $SO(10)$ based models 
with Yukawa coupling unification upon
implementation of the above specific set of scalar mass boundary 
conditions\cite{imh,imh_2}. 
Our results here are obtained using a bottom-up approach, and
indicate that for $\mu >0$, a high degree of Yukawa coupling unification
can {\it only} be obtained using approximately the boundary conditions
Eq. \ref{eq:2}.
The values of the Yukawa couplings obtained at the GUT scale are
$f_t:0.47-0.50$, $f_b:0.35-0.37$ and $f_\tau :0.47-0.52$. Their magnitudes
are only sufficient to generate a small scalar IMH\cite{imh,imh_2}.
For Yukawa unified solutions with $\mu <0$, we found
a weaker correlation for the soft masses around 
$m_{10}= \sqrt{2}m_{16}$\cite{so10}
but we find no correlation for the trilinear parameter.

In Fig. \ref{fig2}, we show values of selected weak scale sparticle
masses generated from the Yukawa unified model. In frame {\it a}), we show
the $m_A\ vs.\ m_h$ plane of Higgs masses. The light scalar $h$
has a mass clustering about the region $m_h\sim 115-130$ GeV, while
$m_A$ ranges between $100-500$ GeV, and is generally lower than
$m_A$ values generated in models with lower values of $\tan\beta$.
Both the $h$ and $A$ (and also the heavy Higgs $H$) may be accessible
to Higgs searches at the Fermilab Tevatron\cite{tevatron}, 
and the low values of $m_A$
may give rise to measureable rates for $B\to \mu^+\mu^-$ 
decay\cite{babu}.

In frame {\it b}), we show the $m_{\tst_1}\ vs.\ m_{\tb_1}$ plane.
We always find $m_{\tb_1}> m_{\tst_1}$, in contrast to
Yukawa unified models with $\mu <0$.  The value of $m_{\tst_1}$
ranges between $100-600$ GeV for $m_{\tb_1}<1$ TeV.

Frame {\it c}) shows the $m_{\tw_1}\ vs.\ m_{\ttau_1}$ plane. 
We note that $m_{\ttau_1}\agt m_{\tw_1}$, while
$m_{\tw_1}\alt 400$ GeV, and
is likely accessible to a linear $e^+e^-$ collider
operating with $\sqrt{s}\sim 800$ GeV\cite{tesla,nlc}.

Finally, in frame {\it d}), we show the $m_{\te_R}\ vs.\ m_{\tu_R}$
plane. In this case, $m_{\te_R}$ and $m_{\tu_R}$ 
typically lie beyond 1 TeV.
Such high mass values
for first and second generation scalars can act to suppress many CP violating
processes via a decoupling solution; they are generally not sufficiently 
heavy to suitably suppress the most dangerous flavor violating processes, 
such as $K-\overline{K}$ mixing\cite{masiero}. A sample sparticle mass 
spectrum is shown in Table 1 for a Yukawa unified model with $\mu >0$.
In this case, $a_\mu =16.6\times 10^{-10}$, within the $2\sigma$ limit 
from E821\cite{e821}; for our other models, $a_\mu$ typically varies
around $(10\pm 6)\times 10^{-10}$.

The cosmological relic density of neutralinos  has been calculated
in Ref. \cite{so10_2} for Yukawa unified models with $\mu <0$.
Little should change by switching to $\mu >0$: the pseudoscalar
$A$ and heavy scalar Higgs $H$ will still have large 
widths of order $20-50$ GeV
due to the large $b$ and $\tau$ Yukawa couplings, and will be light enough
that $\tz_1\tz_1$ annihilation can take place efficiently through
$s$-channel annihilation. In addition, the rates for direct detection
of relic neutralinos will remain large, as in the $\mu <0$ 
case\cite{so10_2}. The rate for $b\to s\gamma$ can be substantially 
different for $\mu>0$ compared to the $\mu <0$ result, and regions
of parameter space certainly exist where this decay rate falls
within experimental limits. Explicit results for the RIMH model
with $\mu >0$ 
have been shown in Ref. \cite{imh_2}. Finally, the value of $a_\mu$
has been calculated in Ref. \cite{bbft} for Yukawa unified models
with $\mu <0$ and for RIMH models with $\mu >0$. Regions of 
model space with an acceptable $a_\mu$ certainly exist for the
$\mu >0$ case. Further results along these lines will be presented 
in a forthcoming publication. 

Finally, we note that Yukawa unified models 
with $\mu >0$ have also been recently reported by 
Blazek {\it et al.}\cite{bdr}. 
These authors use a top-down approach and adopt independent
values for $m_{H_u}$ and $m_{H_d}$ rather than 
$D$-term splitting amongst scalar masses. 
We verify that in this case also 
Yukawa unified solutions can be obtained,
although in our approach these generally occur at the 35-55\% level. 
The solutions typically have
$m_{H_u}\sim \sqrt{2}m_{16}$, with $m_{H_u}<m_{H_d}$, and
$A_0\simeq -2m_{16}$.

It is well known that the three standard model gauge couplings
approximately unify when extrapolated to the scale 
$M_{GUT}\simeq 2\times 10^{16}$ GeV. This may be regarded as
a coincidence, or as evidence for SUSYGUTs. 
Similarly, as a consequence of Yukawa unification, 
the clustering of the SSB parameters about an 
approximate $SO(10)$ inspired fixed point,
if taken seriously, can be considered as evidence for a
supersymmetric $SO(10)$ grand unified model. 

%
\acknowledgments
This research was supported in part by the U.~S. Department of Energy
under contract number DE-FG02-97ER41022.
J.F. was supported by a spanish MEC-FPI grant and by
the European Commission TMR contract HPRN-CT-2000-00148.
%
%

\newpage
%
%

\iftightenlines\else\newpage\fi
\iftightenlines\global\firstfigfalse\fi
\def\dofig#1#2{\epsfxsize=#1\centerline{\epsfbox{#2}}}

\begin{table}
\begin{center}
\caption{Weak scale sparticle masses and parameters (GeV) for an $SO(10)$
case study.}
\bigskip
\begin{tabular}{lc}
\hline
parameter & value  \\
\hline

$m_{16}$ & 1108.3 \\
$m_{10}$ & 1497.7 \\
$M_D   $ & 58.1 \\
$m_{1/2}$ & 325.2 \\
$A_0$ & -2108.0 \\
$\tan\beta$ & 49.9 \\
$f_t(M_{GUT})$ & 0.484 \\
$f_b(M_{GUT})$ & 0.372 \\
$f_\tau (M_{GUT})$ & 0.518 \\
$m_{\tg}$ & 810.9 \\
$m_{\tu_L}$ & 1267.1 \\
$m_{\td_R}$ & 1253.0 \\
$m_{\tst_1}$& 211.3 \\
$m_{\tb_1}$ & 607.7 \\
$m_{\tell_L}$ & 1120.3 \\
$m_{\tell_R}$ & 1115.4 \\
$m_{\tnu_{e}}$ & 1117.5 \\
$m_{\ttau_1}$ & 308.1  \\
$m_{\tnu_{\tau}}$ & 829.1 \\
$m_{\tw_1}$ & 145.5 \\
$m_{\tz_2}$ & 164.6 \\
$m_{\tz_1}$ & 111.6 \\
$m_h      $ & 121.6 \\
$m_A      $ & 200.1 \\
$m_{H^+}  $ & 225.6 \\
$\mu      $ & 165.4 \\
$a_\mu    $ & $16.6\times 10^{-10}$ \\
\hline
\label{tab:cases}
\end{tabular}
\end{center}
\end{table}

\newpage
%

%
\begin{figure}
\dofig{5in}{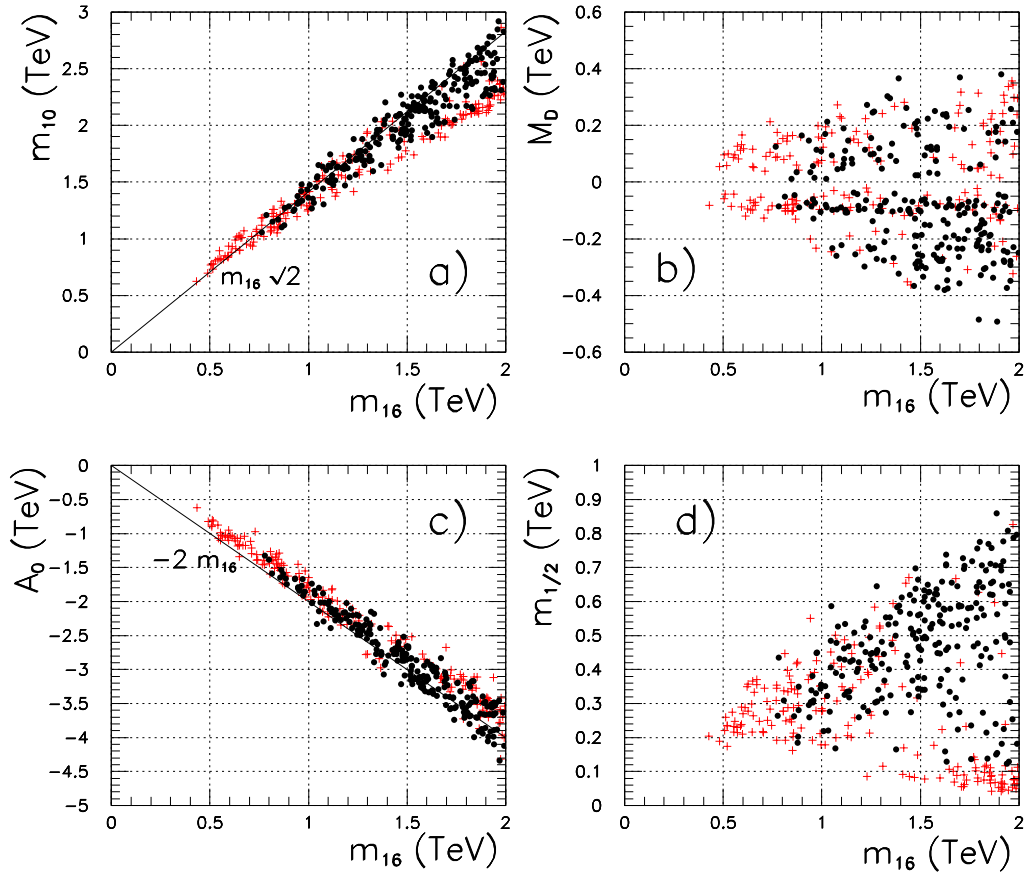}
\caption[]{
Plots of regions of parameter space where valid solutions to 
minimal SUSY $SO(10)$ are obtained, consistent with Yukawa
coupling unification to 45\% with $\mu >0$ and REWSB. Crosses
are excluded by collider searches.}
\label{fig1}
\end{figure}
\begin{figure}
\dofig{5in}{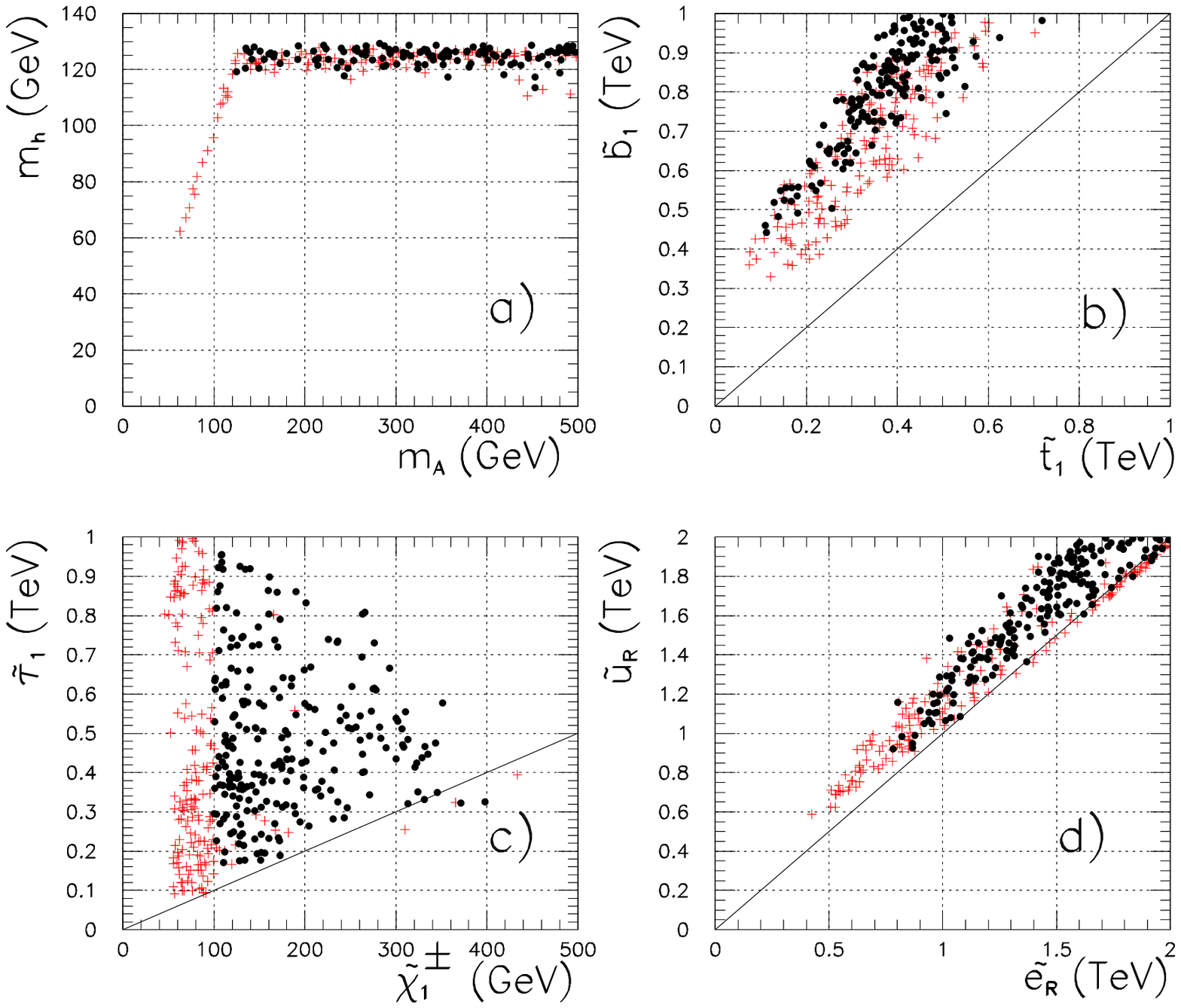}
\caption[]{
The range of selected sparticle masses that are generated in 
minimal SUSY $SO(10)$ models with Yukawa coupling unification to
45\%, $\mu >0$ and REWSB. Crosses are excluded by collider searches.}
\label{fig2}
\end{figure}


\begin{references}
\bibitem{review} H. Georgi, in {\it Proceedings of the American Institue 
of Physics}, edited by C. Carlson (1974); H. Fritzsch and P. Minkowski,
Ann. Phys. {\bf 93}, 193 (1975); M. Gell-Mann, P. Ramond and R. Slansky,
Rev. Mod. Phys. {\bf 50}, 721 (1978). For a recent review, 
see R. Mohapatra, hep-ph/9911272 (1999).
%
\bibitem{seesaw} M. Gell-Mann, P. Ramond and R. Slansky, in {\it Supergravity,
Proceedings of the Workshop}, Stony Brook, NY 1979 (North-Holland, Amsterdam);
T. Yanagida, KEK Report No. 79-18, 1979; R. Mohapatra and G. Senjanovic, 
Phys. Rev. Lett. {\bf 44}, 912 (1980).
%
\bibitem{fmasses}  Some references include
G. Anderson, S. Raby, S. Dimopoulos, L. Hall and G. Starkman,
Phys. Rev. D{\bf 49}, 3660 (1994);
M. Carena, S. Dimopoulos, C. Wagner and S. Raby, 
Phys. Rev. D{\bf 52}, 4133 (1995);
K. Babu, J. Pati and F. Wilczek, Nucl. Phys. {\bf B566}, 33 (2000);
C. Albright and S. Barr, Phys. Rev. D{\bf 58}, 013002 (1998). 
%
\bibitem{isajet} F. Paige, S. Protopopescu, H. Baer and X. Tata,
hep-ph/0001086 (2000).
%
\bibitem{mv} S. Martin and M. Vaughn,  Phys. Rev. D{\bf 50}, 2282 (1994).
%
\bibitem{yuk} L. J. Hall, R. Rattazzi and U. Sarid, 
Phys. Rev. D{\bf 50}, 7048 (1994).
%
\bibitem{bmpz}  D. Pierce, J. Bagger, K. Matchev and R. Zhang,
Nucl. Phys. {\bf B491}, 3 (1997).
%
\bibitem{others} 
M. Carena, M. Olechowski, S. Pokorski and C. Wagner,
Nucl. Phys. B{\bf 426}, 269 (1994); 
B. Ananthanarayan, Q. Shafi and X. Wang, Phys. Rev. D{\bf 50}, 5980 (1994);
R. Rattazzi and U. Sarid, Phys. Rev. D{\bf 53}, 1553 (1996).
%
\bibitem{so10} H. Baer, M. Diaz, J. Ferrandis and X. Tata, 
Phys. Rev. D{\bf 61}, 111701 (2000).
%
\bibitem{so10_2} H. Baer, M. Brhlik, M. Diaz, J. Ferrandis,
P. Mercadante, P. Quintana and X. Tata, 
Phys. Rev. D{\bf 63}, 015007 (2001). 
%
\bibitem{relic} M. Drees and M. Nojiri,  Phys. Rev. D{\bf 47}, 376 (1993);
 H. Baer and M. Brhlik, Phys. Rev. D{\bf 57}, 567 (1998).
%
\bibitem{bsg}  
H. Baer, M. Brhlik, D. Castano and X. Tata, 
Phys. Rev. D{\bf 58}, 015007 (1998).
%
\bibitem{e821} H. N. Brown {\it et al.} (Muon $g-2$ Collaboration),
Phys. Rev. Lett. 86, 2227 (2001).
%
\bibitem{cm} A. Czarnecki and W. Marciano,  
Phys. Rev. D{\bf 64}, 013014 (2001).
%
%
\bibitem{bbft} See {\it e.g.} H. Baer, C. Balazs, J. Ferrandis and X. Tata,
Phys. Rev. D{\bf 64}, 035004 (2001) and references therein.
%
\bibitem{bwright} B. D. Wright, hep-ph/9404217 (1994).
%

\bibitem{Bagger:1995bw}
J.~Bagger, K.~Matchev and D.~Pierce,
Phys.\ Lett.\ B {\bf 348}, 443 (1995)
%
\bibitem{bfpz} 
J. Bagger, J. Feng, N. Polonsky and R. Zhang,
Phys. Lett. {\bf B473}, 264 (2000).
%
\bibitem{imh}  H. Baer, P. Mercadante and X. Tata,
Phys. Lett. {\bf B475}, 289 (2000).
%
\bibitem{imh_2} H. Baer, C. Bal\'azs, M. Brhlik, P. Mercadante,
X. Tata and Y. Wang, Phys. Rev. D{\bf 64}, 015002 (2001).
%
\bibitem{tevatron} M. Carena {\it et al.}, hep-ph/0010338 (2000).
%
\bibitem{babu} K. Babu and C. Kolda, Phys. Rev. Lett. {\bf 84}, 228 (2000).
%
%
\bibitem{tesla} Physics at TESLA, R.D. Heuer {\it et al.}, 
DESY-01-011C (2001). 
%
\bibitem{nlc} T. Abe {\it et al.} (American Linear Collider Working Group),
hep-ex/0106057 (2001).
%
\bibitem{masiero} F. Gabbiani, E. Gabrielli, A. Masiero and L. Silvestrini,
Nucl. Phys. {\bf B477}, 321 (1996).
%
\bibitem{bdr} T. Blazek, R. Dermisek and S. Raby, hep-ph/0107097 (2001).
%
\end{references}
\end{document}